\documentclass[showkeys,aps,prd,nofootinbib,floatfix,amsmath,amssymb]{revtex4}
\usepackage{graphicx}

\begin{document}
\makeatletter
\def\v#1{{\bf#1}}
\def\be{\begin{equation}}
\def\ee{\end{equation}}
\def\bea{\begin{eqnarray}}
\def\eea{\end{eqnarray}}
\def\ahalf{{\textstyle{1\over2}}}
\def\twet{{\textstyle{2\over3}}}
\newcommand{\bfalpha}{\mbox{\boldmath$\alpha$\unboldmath}}
\newcommand{\bfbeta}{\mbox{\boldmath$\beta$\unboldmath}}
\newcommand{\bfxi}{\mbox{\boldmath$\xi$\unboldmath}}
\newcommand{\bfeta}{\mbox{\boldmath$\eta$\unboldmath}}
\newcommand{\bfsigma}{\mbox{\boldmath$\sigma$\unboldmath}}
\newcommand{\bfpi}{\mbox{\boldmath$\pi$\unboldmath}}
\newcommand{\bfgamma}{\mbox{\boldmath$\gamma$\unboldmath}}
\newcommand{\bfrho}{\mbox{\boldmath$\rho$\unboldmath}}
\newcommand{\bfkappa}{\mbox{\boldmath$\kappa$\unboldmath}}
\newcommand{\bfmu}{\mbox{\boldmath$\mu$\unboldmath}}
\newcommand{\sg}{\mbox{\rm sg}}
\def\ie{{\it i.e.\,}}
\def\etal{{\it et al.   }}
\def\ocal{\mbox{$\cal O\,$}}
\def\ycal{\mbox{$\cal Y\,$}}
\def\lcal{\mbox{$\cal L\,$}}
\def\ecal{\mbox{$\cal E\,$}}
\def\hcal{\mbox{$\cal H\,$}}
\def\ncal{\mbox{$\cal N\,$}}
\def\mcal{\mbox{$\cal M\,$}}
\def\fcal{\mbox{$\cal F\,$}}
\def\<{\langle}
\def\>{\rangle}

\newbox\slashbox \setbox\slashbox=\hbox{$/$}
\newbox\Slashbox \setbox\Slashbox=\hbox{\large$/$}
\def\pFMslash#1{\setbox\@tempboxa=\hbox{$#1$}
  \@tempdima=0.5\wd\slashbox \advance\@tempdima 0.5\wd\@tempboxa
  \copy\slashbox \kern-\@tempdima \box\@tempboxa}
\def\pFMSlash#1{\setbox\@tempboxa=\hbox{$#1$}
  \@tempdima=0.5\wd\Slashbox \advance\@tempdima 0.5\wd\@tempboxa
  \copy\Slashbox \kern-\@tempdima \box\@tempboxa}
\def\FMslash{\protect\pFMslash}
\def\FMSlash{\protect\pFMSlash}
\def\miss#1{\ifmmode{/\mkern-11mu #1}\else{${/\mkern-11mu #1}$}\fi}
\makeatother

\title{A classical lower bound on the size of a massive neutrino}
\author{E. Sadurni$^{(a)}$, A. Rosado$^{(a)}$ and S. Rosado-Navarro$^{(b)}$.}
\address{$^{(a)}$Instituto de F\'{\i}sica, Benem\'erita Universidad Aut\'onoma de Puebla.\\
 Apdo. Postal J-48, C.P. 72570 Puebla, Pue., M\'exico.\\
 $^{(b)}$Facultad de Ciencias F\'{\i}sico-Matem\'aticas, Benem\'erita
Universidad Aut\'onoma de Puebla, Pue., Apartado Postal 1364, C.P. 72000, Puebla, Pue., M\'exico.\\
}
\date{\today}
\begin{abstract}
In this paper, we calculate the size of a massive neutrino in the following approach. We perform our calculation using its mass, spin, and magnetic moment through the neutrino-electron interaction, $via$ the classical magnetic dipole-dipole interaction. Thus, our estimate is obtained by mimicking the low-energy electroweak scattering process $\nu_l$-$l^{\prime}$. This leads to surprisingly accurate result which differs in less than one order of magnitude of more detailed calculations with one-loop corrections based on the neutrino charge radius and the $\nu_l$-$l^{\prime}$ scattering process. The resulting estimates are flavour-blind and gauge independent by construction. We also find that our lower bound is below the reported experimental upper bound on the electron neutrino charged radius. So we obtained a constraining range for the neutrino size.
\end{abstract}
\maketitle

\setcounter{footnote}{0} \setcounter{page}{1}
\setcounter{section}{0} \setcounter{subsection}{0}
\setcounter{subsubsection}{0}

\noindent {\bf Classical characteristic length for the neutrino.} 
The classical radius of charge for a neutrino is a very elusive concept. Here, the obvious absence of charge is only the first complication in a series of no-go arguments against such an intrinsic quantity. If the next term in a multipolar field expansion is considered, the particle's magnetic dipole moment $\mu_{\nu}$ may appear initially as a good candidate. However, the classical size of a dipole shows two pitfalls that must be avoided: If a loop of permanent current $J$ is assigned to the dipole, one cannot solve for the loop area $A$ because in such tentative classical model one has $J=0$ and $\mu_{\nu}=J \times A = 0$. On the other hand, if the dipole is thought as a magnetic bar of strength $S$ and length $L$, the formula $\mu_{\nu} = S \times L$ does not allow to solve for $L$, since $S$ is not determined directly from observation (it remains unknown). In addition, this construction entails the use of two opposite magnetic monopoles producing a bar magnet, whose hypothetical existence is beyond known physics. These arguments suggest that the only way to typify a characteristic length for $\nu_l$ is by means of its interaction with any other particle. This procedure works well when particle flavour is irrelevant in the interaction strength. Here we propose the next quantity: a radius $r_{\nu-l}$ of electroweak interaction arising from $\nu_l$-$l^{\prime}$ dipole-dipole potential energy, which is independent of $l=e, \mu, \tau$. The interaction process that we want to mimic by this procedure is indicated in the diagram of fig. \ref{fig:1}. Here we think of dipoles as electromagnetically interacting objects, where the 'weak' part is taken into account by the very small magnetic moment of $\nu$. We are not going to use Fermi's coupling constant $G_F$ in our estimate \cite{fermicoupling}:
\begin{figure}[h]
\begin{center}
\includegraphics[width=6.5cm]{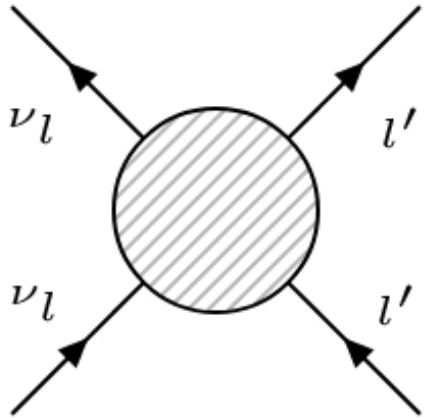} \qquad
\end{center}
\caption{\label{fig:1} Interaction in $\nu_l$-$l^{\prime}$ scattering, modelled in our classical estimate as a dipole-dipole interaction. This gives rise to the neutrino interaction length.}
\end{figure}
We start with the potential energy between two magnetic dipoles with moments $\mu_{\nu}, \mu_{l}$ separated by a distance $r$ 
\begin{eqnarray}
V = \frac{\mu_0}{4\pi} \times \frac{3 ({\bfmu}_{\nu} \cdot \hat r)({\bfmu}_{l} \cdot \hat r)-{\bfmu}_{\nu} \cdot {\bfmu}_{l}}{r^3}
\label{eq:1}
\end{eqnarray}
where $\mu_0$ is the magnetic constant and $\hat{r}$ is the unit relative vector. Here we recall that ${\bfmu} \propto \v S$, but since we are carrying out a classical estimate, we shall ignore for the moment the presence of spin operators (their average will be shown to be in the order of unity later on). Then, the magnitude of (\ref{eq:1}) is controlled by the physical constants

\begin{eqnarray}
|V| \sim \frac{\mu_0}{4\pi} \times \frac{\mu_{l} \mu_{\nu}}{r^3} \equiv W, \quad \mu_l = \gamma_l \mu_{\rm B}, \quad \mu_{\nu} \lesssim 0.29 \times 10^{-10} \mu_{\rm B}
\label{eq:2}
\end{eqnarray}
where $\mu_{\rm B}= e \hbar / 2 m_e c$ is Bohr's magneton and $\gamma_l$ is the gyromagnetic factor containing possible anomalies. As reported in \cite{pdg} (and as discussed in, e.g., \cite{dipole}), we have $\gamma_l-1 \lesssim 10^{-3}$ for $l = e, \mu$ and $\lesssim 10^{-2}$ for $l=\tau$, so this quantity is virtually flavour-blind. To make the potential energy (\ref{eq:1}) gauge independent, we may regard it as the necessary work $W$ to bring the two (point-like) particles to a distance $r$ adiabatically, and when one of them starts at infinity. We conceive this as the low-energy limit of a classical scattering event, and since particle creation is
not allowed in this process, we bound $W$ from above by the lightest rest mass in the particle pair, namely

\begin{eqnarray}
\frac{\mu_0}{4\pi} \times \frac{\mu_{l} \mu_{\nu}}{r_{\nu-l}^3} = m_{\nu} c^2. 
\label{eq:3}
\end{eqnarray}
Now we use the reported value $m_{\nu} c^2 \lesssim 2.1$ eV \cite{pdg} to solve for $r$. In all fairness, the r.h.s of (\ref{eq:3}) does not depend on flavour either, because $m_{\nu}=\< M \>$ must be an average of the mass operator $M$ over flavour states. With this information, the neutrino interaction radius becomes

\begin{eqnarray}
r_{\nu-l} \gtrsim 1.93\times 10^{-16} {\rm cm,\ } \quad r_{\nu-l}^2 \gtrsim 3.72 \times 10^{-32} {\rm cm^2\ } 
\label{eq:4}
\end{eqnarray} 
which is surprisingly close to the more accurate estimates reported by other authors
\cite{lucio,B:2000hf,P:2003rx,B:2004jr,Barranco:2007ea,ValuesEWR}. 
The reported \cite{Deniz:2009mu,pdg} experimental bound is $\<r_{\nu_e}^2\> \times 10^{32} < 3.3$ cm$^2$, so our estimate is slightly above, but very close to the upper limit. In essence, we have shown that the order of magnitude of the radius can be given in classical terms, if we rely on experimental values of magnetic moments and masses. This is remarkable, given the recent improvements in $m_{\nu}$ upper bounds and the potentially undesirable large radii estimates when old values for $m_{\nu}$ are employed. It is concluded that specific details of $r_{\nu-l}$ must be determined by detailed scattering amplitudes with loop corrections and without gauge dependence, as shall be clarified elsewhere. 

We now improve our results in (\ref{eq:4}) by computing the quantum-mechanical average of the operator in the numerator of (\ref{eq:1}). This quantity depends crucially on the state we employ in expectation values, where further refinements can be obtained by additional ensemble averaging with a prescribed density matrix. From the tensor structure of $\mcal \equiv 3 (\bfmu_{\nu} \cdot \hat r)(\bfmu_{l} \cdot \hat r)-\bfmu_{\nu} \cdot \bfmu_{l} $, we see that both spin and orbital parts of the state must be judiciously chosen. Examples of vanishing dipole-dipole averages can be found in the Appendix. Let us choose a state with well defined total angular momentum $\v J= \v L + \v S_{\nu} + \v S_{l}$. The lowest possible state $|j=0,m_j=0\>$ has no projection, which is advantageous in that it yields a trivial average over $m_j$; one can show that the choice $l=s=0$ does give a vanishing average. In the case $l=s=1$ one has the expansion

\begin{eqnarray}
|j=0,m_j=0\> = \frac{1}{\sqrt{3}} \left( |l=1,-1\> |s=1,1\> + |l=1,1\> |s=1,-1\> - |l=1,0\> |s=1,0\>\right). \nonumber \\
\label{eq:5}
\end{eqnarray}

The operator $\mcal $ in question is a scalar, made of two second-rank tensors, therefore we consider now the average of the proportional operator $\ncal = \hbar^2 \mcal / (4 \mu_{\nu}\mu_{l})$. It is easy to show that this operator can be written as

\begin{eqnarray}
\ncal = \left[3 \hat r_i \hat r_j - \delta_{ij}\right]\left[\frac{(\v S_{\nu})_i (\v S_{l})_j + (\v S_{\nu})_j (\v S_{l})_i}{2} - \frac{\delta_{ij}}{3} \v S_{\nu} \v S_{l} \right] = \sum_{q=-2}^{2} T_q^{(2)} S_q^{(2)}.
\label{eq:aa6}
\end{eqnarray}
Using the Wigner-Eckart theorem and a Clebsch-Gordan expansion, yields

\begin{eqnarray}
\<0,0|\ncal|0,0\> &=&\sum_{m_s,m'_s,m_l,m'_l,q} \frac{\< l|| T ||l \> \< s || S || s \>}{\sqrt{(2l+1)(2s+1)}}
\< (11)0,0 |m_l m_s \> \< m_l' m_s' |(11)0,0 \>  \nonumber \\ &\times & \< m_s'q | (12) 1,m_s \> \< m_l' q |(12)1,m_l \> 
\nonumber \\ &=&  \frac{\< l|| T ||l \> \< s || S || s \>}{\sqrt{(2l+1)(2s+1)}} \sum_{m_l=-1}^{1} |\<(11)0,0|m_l\,-m_l \>|^2  \nonumber \\ &\times & \< -m_l\,0|(12)1,-m_l \> \< m_l\,0|(12)1,m_l \> \nonumber \\
&=& \frac{\< 1|| T ||1 \> \< 1 || S || 1 \>}{3} \times \left( \frac{1}{5} \right),
\label{eq:aa7}
\end{eqnarray}
obtained directly from Clebsch-Gordan tables. The reduced matrix elements are evaluated using a specific projection:

\begin{eqnarray}
\< l || T || l \> = \sqrt{2l+1} \frac{\<l0|T_0^{(2)}|l0\>}{\<00|(l2)0,0 \>},
\label{eq:aa8}
\end{eqnarray}
similarly for spin. The corresponding matrix elements of the projections $T_0^{(2)}, S_0^{(2)}$ are obtained via

\begin{eqnarray}
S_0^{(2)} &=& \frac{2 (\v S_{\nu})_3 (\v S_{l})_3 - (\v S_{\nu})_1 (\v S_{l})_1 - (\v S_{\nu})_2 (\v S_{l})_2}{\sqrt{6}} \nonumber \\
T_0^{(2)} &=& \frac{3 \cos^2 \theta -1}{\sqrt{6}}
\label{eq:aa9}
\end{eqnarray}
such that

\begin{eqnarray}
\frac{ \< 1 || T || 1 \>}{\sqrt{3}} \times \frac{\< 1 || S || 1 \>}{\sqrt{3}} = \left(-\frac{4}{\sqrt{15}}\right) \left(-\sqrt{\frac{5}{3}}\right) \frac{\hbar^2}{4} = \frac{4}{3} \times \frac{\hbar^2}{4}
\label{eq:aa10}
\end{eqnarray}

Hence the result for $\<\mcal \>$ in the text, eq(\ref{eq:6})

which leads to

\begin{eqnarray}
\<0,0|\mcal|0,0\> = \frac{4}{15} \times \mu_{\nu} \mu_{l}
\label{eq:6}
\end{eqnarray}
{\it i.e.} a factor $4/15$ smaller than the usual contribution from Bohr's magneton. The final answer becomes

\begin{eqnarray}
r_{\nu-l}^2 \gtrsim 1.54 \times 10^{-32} {\rm cm^2\ } 
\label{eq:7}
\end{eqnarray}
which is within experimental bounds.

{\bf Conclusions.} In this work, we have calculated the classical size for a massive neutrino. 
We have performed our calculation using its mass, spin, and magnetic moment through the neutrino-electron interaction, $via$ the classical magnetic dipole-dipole interaction. Thus, our estimate is obtained by mimicking the low-energy electroweak scattering process $\nu_l$-$l^{\prime}$. This leads to surprisingly accurate result which differs in less than one order of magnitude of more detailed calculations with one-loop corrections based on the neutrino charge radius and the $\nu_l$-$l^{\prime}$ scattering process. We have gotten a lower bound on the neutrino size and to our knowledge, this is the first time that a lower bound for the $\nu$ size is reported. The resulting estimate is flavour-blind and gauge independent by construction.


Finally, if we combine the result given in (\ref{eq:7}) and the experimental upper bound on  below eq (\ref{eq:4}) we find the following range for the $\nu_e$ size:
\begin{eqnarray}
1.54 \times 10^{-32} \text{ cm}^2 \lesssim \langle r^2_{\nu_e} \rangle \lesssim 3.3 \times 10^{-32} \text{ cm}^2,
\label{eq:22}
\end{eqnarray}
which can be rewritten as follows:
\begin{eqnarray}
1.24 \times 10^{-16} \text{cm} \lesssim \langle r_{\nu_e} \rangle \lesssim 1.82 \times 10^{-16} \text{cm.}
\label{eq:23}
\end{eqnarray}
It is remarkable that the current experimental upper bound on the electron-neutrino size reported in Refs.\cite{Barranco:2007ea,Deniz:2009mu} are very close to the value of the $\langle r^2_{\nu} \rangle$ calculated in our approach. 

\begin{center}
{\bf ACKNOWLEDGMENTS}
\end{center}
This work was supported in part by the {\it Sistema Nacional de Investigadores (SNI) de M\'exico}. 

\section*{Appendix: Vanishing averages for low states}
Vanishing dipole-dipole interactions (in average) come from inadequate choices of states.
For instance, a vanishing answer will be obtained if a classical average over $\hat r$ is considered. This alone entails a solid-angle integration $(1/4\pi)\int d\Omega$, which is equivalent to a quantum-mechanical average using the spherically symmetric orbital $|l=0,m_l=0\>$ and arbitrary spin $|\chi\>$; one has

\begin{eqnarray}
\< (\v S_{\nu} \cdot \hat r)(\v S_{l} \cdot \hat r) \> = \begin{cases} \frac{\hbar^2}{3 \times 4} (\hat n_{\nu} \cdot \hat n_{l}), & \mbox{if}\quad |\chi \> = |\v S \cdot \hat n_{\nu},+ \>|\v S \cdot \hat n_{l},+ \> \quad \mbox{(product)}\\ \frac{\hbar^2}{3 \times 4}, & \mbox{if}\quad |\chi \> = |s=0,m_s=0 \> \quad \mbox{(singlet)} \end{cases}
\label{eq:a5}
\end{eqnarray}
and
\begin{eqnarray}
\< \v S_{\nu} \cdot \v S_{l} \> = \begin{cases} \frac{\hbar^2}{4} (\hat n_{\nu} \cdot \hat n_{l}), & \mbox{if}\quad |\chi \> = |\v S \cdot \hat n_{\nu},+ \>|\v S \cdot \hat n_{l},+ \> \quad \mbox{(product)}\\ \frac{\hbar^2}{4}, & \mbox{if}\quad |\chi \> = |s=0,m_s=0 \> \quad \mbox{(singlet)} \end{cases}
\label{eq:a6}
\end{eqnarray}
Hence $\< \mcal \>_{l=0}=0$. In the language of spherical tensors, this means that the lowest states producing non-vanishing averages must have at least vector structure $l=1$, {\it i.e.} P orbitals. Here the total spin is also important: To simplify our argument, let us put $|\chi \>$ as a spin singlet $s=0$ to find that the totally coupled state is just the product
\begin{eqnarray}
|(ls),j=1,m_j\> = | l=1, m_l \> \times | s=0, m_s=0 \>.
\label{eq:a7} 
\end{eqnarray}
Then, direct integration (or Wigner-Eckart theorem) and projection averaging $(1/3)\sum_{m_l}$ leads to

\begin{eqnarray}
\< (\v S_{\nu} \cdot \hat r)(\v S_{l} \cdot \hat r) \> &=& \frac{1}{3}\sum_{m_l=-1,0,1} \sum_{i,j} \int d\Omega \big| Y_1^{m_l} \big|^2 \hat r_i \hat r_j \<00| (S_{\nu})_i (S_{l})_j |00\> \nonumber \\ 
&=& - \frac{\hbar^2}{4} \times \frac{1}{3}\sum_{m_l=-1,0,1} \int d\Omega \big| Y_1^{m_l} \big|^2 \left[ \cos^2\theta + \sin^2 \theta (\cos^2 \phi + \sin^2 \phi) \right] \nonumber \\
&=& - \frac{\hbar^2}{4},
\label{eq:a8}
\end{eqnarray}
meanwhile, the spin-spin average yields

\begin{eqnarray}
\< \v S_{\nu} \cdot \v S_{l} \>_{s=0} = \frac{\hbar^2}{2} (s(s+1)-3/4-3/4) = -\frac{3\hbar^2}{4}
\label{eq:a9}
\end{eqnarray}
and with this $\< \mcal \>_{l=1,s=0}= 3(-\hbar^2/4)-(-3\hbar^2/4)=0$ again. From here it is evident that states with $l=s=1$ are needed for non-vanishing results. However, product states of the form (orbital $\times$ spinor) also give 0, after computing the average over orbital projections $(1/3)\sum_{m_l=-1,0,1}$: First we have $|\psi \> = |l=1,m_l\> \times |s=1,m_s\>$. For parallel spins $|s=1,m_s=1\>$ and $m_l=0$ one has

\begin{eqnarray}
\< 3(\hat r \cdot \v S_{\mu})(\hat r \cdot \v S_{l})- \v S_{\mu} \cdot \v S_{l} \> = 3 \left( \frac{3}{5} \frac{\hbar^2}{4} \right) - \frac{\hbar^2}{4} = \frac{4}{5} \frac{\hbar^2}{4}
\label{eq:a10}
\end{eqnarray}
similarly, for $m_l= \pm 1$ and the same spin state
\begin{eqnarray}
\< 3(\hat r \cdot \v S_{\mu})(\hat r \cdot \v S_{l})- \v S_{\mu} \cdot \v S_{l} \> = 3 \left( \frac{3}{15} \frac{\hbar^2}{4} \right) - \frac{\hbar^2}{4} = -\frac{2}{5} \frac{\hbar^2}{4}
\label{eq:a10.1}
\end{eqnarray}
Here we see that the overall average $(1/3)(4/5-2/5-2/5) \hbar^2/4 = 0$, so $(1/3)\sum_{m_l} \<\mcal \>_{l=s=1} = 0$. The result is similar for both spins pointing downwards, {\it i.e.} $|s=1,m_s=-1\>$. In general, orbital projection averages vanish for product states (orbital $\times$ spinor):

\begin{eqnarray}
\< \mcal \> &=& \frac{1}{2l+1} \sum_{m_l=-1,0,1} \int d\Omega \big|Y_l^{m_l} \big|^2 (3 \hat r_i \hat r_j - \delta_{ij}) 
\< (\bfmu_{\nu})_i (\bfmu_{l})_j \>_{s,m_s} \nonumber \\ &=&\frac{1}{4\pi} \int d\Omega (3 \hat r_i \hat r_j - \delta_{ij}) 
\< (\bfmu_{\nu})_i (\bfmu_{l})_j \>_{s,m_s} = 0
\label{eq:a11}
\end{eqnarray}
which is obtained by direct integration of $3 \hat r_i \hat r_j - \delta_{ij}$. Here, off-diagonal $i \neq j$ elements disappear after azimuth integration, while diagonal elements disappear after polar integration.

Finally, we consider states with well defined total angular momentum:
\begin{eqnarray}
|(ls)j,m_j\> = \sum_{m_l,m_s} \< m_l m_s |(ls) j,m_j \> |l,m_l\> |s,m_s\>.
\label{44}
\end{eqnarray}
If an additional average over both $m_j=-j,...,j$ and $|l-s|<j<l+s$ is considered, we immediately see, using the orthogonality of Clebsch-Gordan coefficients, that

\begin{eqnarray}
\<\mcal\> &= &\sum_{j=|l-s|}^{l+s}\sum_{m_j=-j}^{+j}\, \sum_{m_l,m_l',m_s,m_s'} \<(ls)j,m_j|m_l m_s \> \< m_l m_s |(ls)j,m_j \> \nonumber \\ &\times & \int d\Omega Y_l^{m_l *} Y_l^{m_l'}(3\hat r_i \hat r_j - \delta_{ij}) \< s m_s|(\bfmu_{\mu})_i (\bfmu_{l})_j|s m_s' \> \nonumber \\ &=& \sum_{m_l, m_s}\int d\Omega \big| Y_l^{m_l} \big|^2(3\hat r_i \hat r_j - \delta_{ij}) \< s m_s|(\bfmu_{\mu})_i (\bfmu_{l})_j|s m_s \> 
\label{55}
\end{eqnarray}
which falls into a previous case after using the Uns\"old summation theorem for spherical harmonics. 

For the reasons above, we must consider $j$ fixed, {\it i.e.} well defined total angular in a scattering process. 


\begin{references}
\bibitem{ValuesEWR}
We updated the values given by eq.(A5) in Ref\cite{lucio} for the electroweak radius and taking the values for the SM parameters as given in  Ref.\cite{pdg}, we found that:
$\langle r_{\nu_l} \rangle_{EW} = 6.53, \, 5.07, \, 4.13\times 
10^{-17} \text{cm}$ for $l=e, \, \mu \, \tau$, respectively.
\end{references}
\end{document}